\documentclass[11pt]{JHEP3} 
\usepackage{amsfonts,ulem,graphicx,wrapfig,epsfig,bbold}
\usepackage{amsmath,amssymb}

\newcommand{\om}{\Omega_{m}}

\newcommand{\ola}{\Omega_{\Lambda}}

\newcommand{\be}{\begin{equation}}
\newcommand{\ee}{\end{equation}}
\newcommand{\bea}{\begin{eqnarray}}
\newcommand{\eea}{\end{eqnarray}}
\newcommand{\p}{\partial}
\newcommand{\nn}{\nonumber}

\newcommand{\gmn}{g_{\mu\nu}}

\def\Lag{\mathcal{L}}

\def\td{\mathrm{d}}
\def\ph{\phantom}
\title{Cosmological Solutions in Bimetric Gravity and their Observational
  Tests  }

\author{Mikael von Strauss\footnote{\tt mvs@fysik.su.se}~,
Angnis Schmidt-May\footnote{\tt angnis.schmidt-may@fysik.su.se}~,
Jonas Enander\footnote{\tt enander@fysik.su.se}~,
Edvard   M\"ortsell\footnote{\tt edvard@fysik.su.se}~,
S.~F.~Hassan\footnote{\tt fawad@fysik.su.se}
\\  
	Department of Physics \& 
        The Oskar Klein Centre for Cosmoparticle Physics,\\
        Stockholm University, AlbaNova University Centre, 
        SE-106 91 Stockholm, Sweden }

\author{\\ \\}

\abstract{We obtain the general cosmological evolution equations for a
  classically consistent theory of bimetric gravity. Their analytic
  solutions are demonstrated to generically allow for a cosmic
  evolution starting out from a matter dominated FLRW universe and
  relaxing towards a de Sitter (anti-de Sitter) phase at late cosmic
  time. In particular, we examine a subclass of models which contain
  solutions that are able to reproduce the expansion history of the
  cosmic concordance model inspite of the nonlinear couplings of the
  two metrics. This is demonstrated explicitly by fitting these models
  to observational data from Type Ia supernovae, Cosmic Microwave
  Background and Baryon Acoustic Oscillations. In the appendix we
  comment on the relation to massive gravity.}


\preprint{}

\begin{document} 

\section{Introduction and summary}

Cosmological observations confirm to a high degree of precision that
our universe is homogeneous and isotropic and can be described by a
single metric using Einstein's theory of general relativity with the
addition of a cosmological constant $\Lambda_{\mathrm{obs}}$. At the
same time, one of the corner stones of this theory is the equivalence
principle which, taken together with quantum field theory, provides
one of the most serious unresolved problems of modern theoretical
physics. Quantum field theory suggests the reality of vacuum energy
and naturally provides a cosmological constant
$\Lambda_{\mathrm{vac}}$, while the equivalence principle guarantees
that this energy gravitates with the same strength as any other
source. The main problem is to reconcile the expected large value of
$\Lambda_{\mathrm{vac}}$ with the observed small value of
$\Lambda_{\mathrm{obs}}$, without recourse to extreme fine tuning. This
is one of the main motivations for considering theories of modified
gravity.

An intuitive modification of general relativity is massive gravity
which generically also suffers from a ghost instability
\cite{Boulware:1973my}. Such theories must contain at least two metrics, say,
$g_{\mu\nu}$ and $f_{\mu\nu}$, where $f_{\mu\nu}$ is treated as fixed
and non-dynamical. This extra metric is needed to construct
non-derivative, non-linear mass terms in the action; since $\sqrt{-g}$
and $g^\mu_{\ph\mu\mu}=4$ alone are not adequate for the purpose.
Massive gravity has received increased attention since a class of
massive actions, formulated with a flat $f_{\mu\nu}$, was proposed in
\cite{deRham:2010ik,deRham:2010kj} and were shown to be ghost free at the
completely non-linear level\cite{Hassan:2011hr}. These were soon
generalized to arbitrary $f_{\mu\nu}$ and proved to remain ghost free
\cite{Hassan:2011tf,Hassan:2011ea}. The potential relevance of massive
gravity to the cosmological constant problem can be argued on a
heuristic level, based on the Yukawa suppression of the massive
amplitudes over large distances. This may be expected to weaken
gravity over such distances and mimic cosmic acceleration or possibly
screen out a large cosmological constant. ``Self-accelerated''
solutions of massive gravity with a small cosmological constant have
been considered, for example, in
\cite{deRham:2010tw,Gumrukcuoglu:2011ew}. However, the screaning
mechanism does not seem to work without fine tuning
\cite{Hassan:2011vm}.

Here, we will consider a generalization of massive gravity, the
bimetric gravity, where $f_{\mu\nu}$ is promoted to a dynamical
variable with its own kinetic term.  The idea of studying bimetric
gravity for cosmological purposes has a long history
\cite{Rosen:1940zz,Rosen:1975kk} and such theories are known to admit
cosmological vacuum solutions \cite{Salam:1976as}. For more recent
work on bimetric theories see
\cite{ArkaniHamed:2002sp,Banados:2008fi,Berezhiani:2007zf,Blas:2009my,
  Boulanger:2000rq,Damour:2002ws,Milgrom:2009gv} and references
therein. In these studies the nonlinear bimetric interaction potential
is often chosen with the only restriction that it reduces to the
Fierz-Pauli form \cite{Fierz:1939ix,Pauli:1939xp} at the linearized
level. However, these theories also contain the Boulware-Deser ghost
\cite{Boulware:1973my}, similar to massive gravity, and hence are not
consistent. It has been known that internal consistency imposes
restrictions on the kinetic structure of bi-metric
theories \cite{Aragone:1971kh}. Recently, a class of ghost free
bimetric theories were constructed in \cite{Hassan:2011zd}. Here we
will focus on the cosmological implications of these
theories. Spherically symmetric vacuum solutions of these models were
considered in \cite{Comelli:2011wq}.

Bimetric theories describe a pair of interacting massless and massive
spin-2 fields \cite{Salam:1976as,Isham:1971gm} which are combinations of
the two metrics. The presence of the massless mode distinguishes
bimetric gravity from pure massive gravity.  A priori it is not
obvious what combination of the spin-2 fields should couple to the
matter, except that the coupling should not reintroduce the ghost. In
this paper we consider the bimetric theory of \cite{Hassan:2011zd} and
assume that only one of the metrics, $\gmn$, is coupled to
matter. Then, although not a mass eigenstate, this metric alone will
determine the geodesics and the causal structure of the spacetime.

Looking for cosmological solutions we take both of the metrics to be
homogeneous and isotropic and obtain the general cosmological
equations for the metric $g_{\mu\nu}$. Solving these in their most
general form involves finding the zero locus of a quartic
polynominal. This guarantees that solutions of the general equations
exist and they are straightforward to derive. From the general
equation we conclude some generic properties of the solution inherent
to the model. First, they generically asymptote towards a de Sitter or
anti-de Sitter (AdS) spacetime and hence will in general be able to
mimic late-time cosmic acceleration. Secondly, one branch of the
solutions always allows for an expansion history which starts out from
an ordinary FLRW like universe at high densities, and hence the usual
early universe considerations still apply for this solution.

Displaying the full solutions of the quartic in their analytic form is
not very illuminating for general values of the parameters of the
theory. We study instead the analytic details of two simpler classes
of solutions, where the quartic is reduced to a quadratic. These
solutions are shown to exhibit the general behaviour expected of the
full solution. In particular, we identify and study a ``minimal''
bimetric model in more detail. In this model, the Friedman equation is
shown to be completely degenerate with general relativity, up to a
possible rescaling of $G_N$, the Newton's constant. Hence,
observationally it is indistinguishable from the usual concordance
model on cosmological scales. We demonstrate this explicitly by
fitting the model to observations, using data from Type Ia supernovae
(SNe Ia), Cosmic Microwave Background (CMB) and Baryon Acoustic
Oscillations (BAO). We further study neighbouring models of the
minimal model and find that cosmological data favours models close the
minimal one.

The paper is organized as follows: In section \ref{bigeom} we review
ghost free bi-metric gravity and present the equations of motion for a
homogeneous and isotropic ansatz for both metrics. In section
\ref{solutions} we discuss the general properties of cosmological
solutions and then focuses on two more constrained models that allow
for solutions close to those of general relativity. We specify the
model used for the comparison to observations in section
\ref{sec_parameter} and summarize the cosmological data considered in
section \ref{data}. Section \ref{results} contains the results of the
model fit, which are discussed in section \ref{discussion}.
Appendix \ref{app_eom} contains the details for the equations of
motion in our metric ansatz. In Appendix \ref{app_massgrav} we
discuss the Bianchi constraints and comment on the relation to some
cosmological solutions recently obtained in the massive gravity
literature.

\section{The bimetric gravity equations in the cosmological ansatz}
\label{bigeom}

In this section we review the structure of ghost free bimetric gravity
and its equations of motion. We then write the equations of motion
using a cosmological ansatz for the metrics.

\subsection{Review of bimetric gravity action and equations of motion}
\label{review}
The action for the metric $g_{\mu\nu}$ interacting with another spin-2
field $f_{\mu\nu}$ through a non-derivative potential is determined by
the requirement of the absence of the Boulware-Deser ghost. The most
general action of this type, modulo choice of coupling to matter, has
the form  
\cite{Hassan:2011zd},
\begin{align}
S= & -\frac{M_g^2}{2}\int\td^4x\,\sqrt{-\det g}~R(g)
-\frac{M_f^2}{2}\int\td^4x\,\sqrt{-\det f}~R(f)
\label{action} \\
 & + m^2M_g^2\int\td^4x\sqrt{-\det g}
\sum_{n=0}^4\beta_n\,e_n(\sqrt{g^{-1}f}) + \int\td^4x\sqrt{-\det g}~
\Lag_m(g,\Phi)\nn \,.
\end{align}
Here, $\beta_n$ are free parameters and $e_n(\mathbb{X})$ are
elementary symmetric polynomials of the eigenvalues of the matrix
$\mathbb{X}$ given explicitly by  
\begin{align}
	e_{0}(\mathbb{X}) &= 1\,,\hspace{1.4cm}
	e_{1}(\mathbb{X}) = [\mathbb{X}]\,, \hspace{1.3cm}
	e_{2}(\mathbb{X}) = \tfrac1{2}([\mathbb{X}]^2-[\mathbb{X}^2])\,,
\nn\\[.1cm]
	e_{3}(\mathbb{X}) &= \tfrac1{6}([\mathbb{X}]^3
		-3[\mathbb{X}][\mathbb{X}^2]+2[\mathbb{X}^3])\,,\qquad  
	e_{4}(\mathbb{X}) = \det(\mathbb{X})\,,
\end{align}
where the square brackets denote the matrix trace. The non-trivial
point here is the appearance of the square root matrix
$\mathbb{X}=\sqrt{g^{-1}f}$ which is necessary to avoid the ghost at
the nonlinear level.

As demonstrated through a nonlinear ADM analysis in
\cite{Hassan:2011ea,Hassan:2011zd}, this action is ghost free and contains 7
propagating degrees of freedom; 2 corresponding to a massless spin-2
graviton and 5 corresponding to a massive spin-2 field. Both
$g_{\mu\nu}$ and $f_{\mu\nu}$ are combinations of these massless and
massive degrees of freedom.

Since $e_n\sim (\sqrt{g^{-1}f})^n$, the $\beta_n$ parameterize the
order of interactions at the nonlinear level. As such they are more
convenient to use than other parameterizations. Besides the two Planck
masses, the action (\ref{action}) contains five free parameters
$\beta_n$. Of these, $\beta_0$ and $\beta_4$ parameterize the
cosmological contants of $g_{\mu\nu}$ and $f_{\mu\nu}$, respectively. 
One combination of the remaining $ \beta$'s gives the mass of the
massive mode, leaving us with two extra free parameters to
characterize the nonlinear interactions.

For convenience, and also not to violate the equivalence principle in
any drastic way, we only couple $g_{\mu\nu}$ to matter. This metric
determines the geodesics and the causal structure of spacetime.
Apart from this coupling, the action is invariant under the exchange, 
\be
g \leftrightarrow f\,,\qquad 
\beta_n\rightarrow \beta_{4-n}\,,\qquad
M_g\leftrightarrow M_f\,,\qquad 
m^2\rightarrow m^2 M_g^2/M_f^2
\label{f-g}
\ee
where the last replacement is needed due to our asymmetric
parameterization of the mass scale in terms of $m^2$.

Setting $\beta_3=0$ in the action (\ref{action}) eliminates the
highest order interaction term in $\sqrt{g^{-1}f}$. However, in view
of (\ref{f-g}) we still have a cubic order interaction term in
$\sqrt{f^{-1}g}$ which can in turn be eliminated by setting
$\beta_1=0$. In this sense, the choice $\beta_1=\beta_3=0$ leads to
the ``minimal'' bimetric action, which is the simplest in the class. 

Let us now consider the equations of motion and the Bianchi
constraints that follow from (\ref{action}). Varying the action with
respect to $g_{\mu\nu}$ gives the equations of motion
\cite{Hassan:2011vm}, 
\be
\label{g_eom}
R_{\mu\nu}-\tfrac{1}{2}\gmn R + \frac{m^2}{2}\sum_{n=0}^3(-1)^n\beta_n 
\left[g_{\mu\lambda}Y_{(n)\nu}^\lambda(\sqrt{g^{-1}f})+g_{\nu\lambda}
Y_{(n)\mu}^\lambda(\sqrt{g^{-1}f})\right]=\tfrac{1}{M_g^2}T_{\mu\nu}\,,
\ee
where $Y_{(n)\nu}^\lambda(\sqrt{g^{-1}f})
$ are defined below.
Similarly, varying with respect to $f_{\mu\nu}$ gives, 
\be
\label{f_eom}
\bar{R}_{\mu\nu}-\tfrac{1}{2}f_{\mu\nu} \bar{R} +\frac{m^2}{2M_\star^2} 
\sum_{n=0}^3(-1)^n\beta_{4-n}\left[f_{\mu\lambda}
Y_{(n)\nu}^\lambda(\sqrt{f^{-1}g})+f_{\nu\lambda} 
Y_{(n)\mu}^\lambda(\sqrt{f^{-1}g}) \right]=0\,, 
\ee  
where the overbar indicate $f_{\mu\nu}$ curvatures and we have
introduced the dimensionless ratio of Planck masses 
\be
M_\star^2\equiv\frac{M_f^2}{M_g^2}\,.
\ee
Note that (\ref{f_eom}) is essentially obtainable from (\ref{g_eom})
through the replacements (\ref{f-g}). Finally, the matrices
$Y_{(n)\mu}^\lambda(\mathbb{X})$ in (\ref{g_eom}) and (\ref{f_eom}) are 
given by (with square brackets denoting the
trace)\cite{Hassan:2011vm},    
\begin{align}
Y_{(0)}(\mathbb{X}) &= \mathbb{1}\,, \quad
Y_{(1)}(\mathbb{X}) = \mathbb{X}-\mathbb{1}[\mathbb{X}]\,,\nn\\[.1cm]
Y_{(2)}(\mathbb{X}) &= \mathbb{X}^2
-\mathbb{X}[\mathbb{X}]+\tfrac{1}{2}\mathbb{1}([\mathbb{X}]^2-
[\mathbb{X}^2])\,,\nn\\[.1cm]
Y_{(3)}(\mathbb{X}) &= \mathbb{X}^3-\mathbb{X}^2[\mathbb{X}]
+\tfrac{1}{2}\mathbb{X}([\mathbb{X}]^2-[\mathbb{X}^2])
-\tfrac{1}{6}\mathbb{1}([\mathbb{X}]^3-3[\mathbb{X}]
[\mathbb{X}^2]+2[\mathbb{X}^3])\,.
\end{align}

As a consequence of the Bianchi identity and the covariant
conservation of $T_{\mu\nu}$, the $g_{\mu\nu}$ equation of motion
(\ref{g_eom}) leads to the Bianchi constraint,
\be 
\label{bianchi_g}
\nabla^\mu\sum_{n=0}^3(-1)^n\beta_n \left[
g_{\mu\lambda}Y_{(n)\nu}^\lambda(\sqrt{g^{-1}f}) +
g_{\nu\lambda}Y_{(n)\mu}^\lambda(\sqrt{g^{-1}f})\right]=0.
\ee
Similarly, the $f_{\mu\nu}$ equation of motion (\ref{f_eom}) leads to
the Bianchi constraint,
\be
\bar{\nabla}^\mu\sum_{n=0}^3(-1)^n\beta_{4-n}\left[
f_{\mu\lambda}Y_{(n)\nu}^\lambda(\sqrt{f^{-1}g}) +
f_{\nu\lambda}Y_{(n)\mu}^\lambda(\sqrt{f^{-1}g})\right]=0.
\label{bianchi_f}
\ee
where the overbar indicates covariant derivatives with respect to
the $f_{\mu\nu}$ metric. Both of these Bianchi constraints follow from
the invariance of the interaction term under the diagonal subgroup of
the general coordinate transformations of the two metrics and hence
are equivalent. From now on we explicitly concentrate on
(\ref{bianchi_g}).  

Before proceeding further, let us briefly remark on the relation to
massive gravity. This corresponds to freezing the dynamics of
$f_{\mu\nu}$ and hence loosing the corresponding equation of motion
(\ref{f_eom}). Hence in the bimetric theory $f_{\mu\nu}$ is much more
constrained than in massive gravity. 

\subsection{Equations in the cosmological ansatz}

We look for solutions where both of the metrics exhibit spatial
isotropy and homogeneity. For simplicity, we also assume both metrics to
have the same spatial curvature $k=0,\pm1$. Then, modulo time
reparameterizations, the most general form for the metrics is, 
\footnote{Modulo non-perturbative solutions that can exist for certain
  values of the $\beta_n$ parameters \cite{Damour:2002wu} (see also
  \cite{Volkov:2011an}).}  
\be
\begin{array}{l}
g_{\mu\nu}\td x^\mu\td x^\nu = -\td t^2 + a^2(t)\,\td \vec x^2
\\[.2cm]
f_{\mu\nu}\td x^\mu\td x^\nu = -X^2(t)\,\td t^2 + Y^2(t)\,\td \vec x^2
\end{array}
\label{ansatz}
\ee
where
\be
\td\vec x^2 = \frac{\td r^2}{1-kr^2} + r^2\left(
\td\theta^2+\sin^2\theta\td\phi^2\right)\,.
\ee
Obviously, we have used time reparameterizations to set $g_{00}=-1$ so
that the $g_{\mu\nu}$ metric is in the usual FLRW form. No other 
transformations are available to further reduce the number of
functions in the metric ansatz. Thus we need to keep three arbitrary
functions to describe the metrics with the specified symmetries. 
Below we will see that the Bianchi constraint (implied by general
covariance) enforces $X=\frac{\dot Y}{\dot a}=\frac{\td Y}{\td a}$,
thus leaving only two free functions to work with. 

Now, we write down the Bianchi constraint and the equations of motion
for the above parameterization of the metrics. For the ansatz
(\ref{ansatz}) the Bianchi constraint (\ref{bianchi_g}), or
(\ref{bianchi_f}), gives, 
\bea
\label{bianchi_constraint}
\frac{3m^2}{a}\left[\beta_1 +2\frac{Y}{a}\beta_2+\frac{Y^2}{a^2}
\beta_3\right]\left(\dot{Y}-\dot{a}X\right)=0.
\eea
One way this can be satisfied is by setting to zero the expression
within the square brackets. This implies solutions where $Y\propto a$,
with special values for the constant of proportionality. As will be
evident from the Friedmann equation (\ref{Fg}) below, this 
leads to the ordinary general relativistic equations with a
cosmological constant of order $m^2$. Further, the special values
imply a vanishing mass for the massive spin-2 field (as is discussed
further in the Appendix \ref{app_massgrav}). One can also 
check that linear metric perturbations around these backgrounds are 
indistinguishable from general relativity. Hence one concludes that 
this is effectively a decoupled class of solutions that do not modify
general relativity on any scale.

The true dynamical constraint is enforced by the vanishing of the
expression within the round brackets,  
\be 
\label{bianchi_constraint_dyn}
X = \frac{\dot Y}{\dot a} = \frac{\td Y}{\td a}\,. 
\ee
Using this result together with the ansatz (\ref{ansatz}), the
$g_{\mu\nu}$ equations of motion (\ref{g_eom}) lead to the modified
Friedmann equation (for details see Appendix \ref{app_eom}), 
\be
-3\left(\frac{\dot a}{a}\right)^2 -3\frac{k}{a^2} + m^2\left[\beta_0
+ 3\beta_1\frac{Y}{a} + 3\beta_2\frac{Y^2}{a^2}
+\beta_3\frac{Y^3}{a^3}\right] = \frac1{M_g^2}T^0_{\ph00} \,,
\label{Fg}
\ee
and the acceleration equation,
\begin{align}
&-2\frac{\ddot a}{a} - \left(\frac{\dot a}{a}\right)^2 -\frac{k}{a^2}
+ m^2\biggl[\beta_0 +2\beta_1\left(\frac{Y}{a} + \frac{\dot Y}{\dot a}
\right)  \nn\\
&\hspace{3.6cm}+\beta_2\left(\frac{Y^2}{a^2}+2
\frac{Y \dot Y}{a \dot a} \right)
+\beta_3\frac{Y^2 \dot Y}{a^2 \dot a}\biggr] = \frac1{M_g^2}T^1_{\ph11}\,.
\label{Accg}
\end{align}
Here we have used $T^1_{\ph11}=T^2_{\ph22}=T^3_{\ph33}$, consistent
with the symmetries of the spacetime. These obviously reduce to the
ordinary Friedmann and acceleration equations of cosmology in the
limit $m^2\rightarrow 0$, although the solutions will not always be
well defined in this limit. Similarly, the $f_{\mu\nu}$ equations of
motion (\ref{f_eom}) lead to, 
\be
-3\left(\frac{\dot a}{Y}\right)^2 -3\frac{k}{Y^2}
+ \frac{m^2}{M_\star^2}\left(\beta_4
+ 3\beta_3\frac{a}{Y} 
+ 3\beta_2\frac{a^2}{Y^2}
+\beta_1\frac{a^3}{Y^3}\right)  = 0 \,,
\label{Ff}
\ee
and
\be
-2\frac{\dot a\ddot a}{Y \dot Y} - \left(\frac{\dot a}{Y}\right)^2
-\frac{k}{Y^2}+ \frac{m^2}{M_\star^2}\biggl[ \beta_4
+\beta_3\left(2\frac{a}{Y} 
+ \frac{\dot a}{\dot Y}\right)
+\beta_2\left(\frac{a^2}{Y^2}
+2\frac{a \dot a}{Y \dot Y}\right)
+\beta_1\frac{a^2 \dot a}{Y^2 \dot Y}\biggr]  = 0\,.
\label{Accf}
\ee
The first of these, the $f$-Friedmann equation (\ref{Ff}), is in
general a cubic equation for $Y$ so the system can be solved exactly.
Further, as a consequence of already having imposed the Bianchi
constraint the two equations are not independent. Indeed, the
$f$-acceleration equation (\ref{Accf}) is obtained by acting with
$(3+(Y/\dot Y)\p_t)/3$ on the $f$-Friedmann equation. Thus we only need
to consider (\ref{Ff}).   

Let us briefly comment on the source structure. In what follows we
will assume a perfect fluid source $T_{\mu\nu}=(\rho+P)u_\mu u_\nu+
 Pg_{\mu\nu}$ so that in the rest frame,
\be
T^0_{\ph00} = -\rho\,,\quad T^i_{\ph ii} = P\quad
{\mathrm{(no\ sum\ implied)}}\,.
\ee
Assuming also an equation of state of the usual form, $P(t)=w\rho(t)$,
the continuity equation, 
\be
\frac{3}{M_g^2}\frac{\dot a}{a}
\left(P+\rho+\frac1{3}\frac{a}{\dot a}\dot\rho\right) = 0\,,
\ee
tells us that, for $w\neq-1$,
\be
\rho = \rho_0 \left(\frac{a}{a_0}\right)^{-3(1+w)}\,,\quad\mathrm{and}\quad
H = -\frac{\dot\rho/\rho}{3(1+w)}\,.
\label{rho}
\ee
Here $\rho_0$ is the present day energy density. Now, in an expanding universe any source with
$w>-1$ will get diluted as the scale factor grows. Hence, $\dot\rho<0$
and $H>0$.  

\section{Viable cosmological solutions}\label{solutions}

In this section we will consider cosmological solutions in bimetric
gravity. In particular we concentrate on parameter values for which the
solutions are close to cosmological solutions in general relativity
and hence are not observationally ruled out.  

\subsection{General features of solutions}\label{general}
In order to make the analysis more transparent, we define the
dimensionless combinations 
\be
\Upsilon\equiv\frac{Y}{a}\,,\quad
\rho_\star\equiv\rho/3M_g^2m^2\,.
\ee
In terms of these, the two Friedmann equations (\ref{Fg}) and
(\ref{Ff}) can be written as 
\be\label{FgUpsilon}
\frac{\beta_3}{3}\Upsilon^3 + \beta_2\Upsilon^2 + \beta_1\Upsilon
+ \frac{\beta_0}{3}  + \rho_\star - \frac{H^2}{m^2} -\frac{k}{m^2 a^2}
= 0 
\ee
and
\be\label{FfUpsilon}
\frac{\beta_4}{3M_\star^2}\Upsilon^2 + \frac{\beta_3}{M_\star^2}
\Upsilon + \frac{\beta_1}{3M_\star^2}\frac{1}{\Upsilon}
+ \frac{\beta_2}{M_\star^2} -\frac{H^2}{m^2}-\frac{k}{m^2a^2}=0\,.
\ee
Subtracting these two equations to eliminate $H^2$ gives in general a
quartic equation for $\Upsilon$,
\be\label{quartic}
\frac{\beta_3}{3}\Upsilon^4
+\left(\beta_2-\frac{\beta_4}{3M_\star^2}\right)\Upsilon^3
+\left(\beta_1-\frac{\beta_3}{M_\star^2}\right)\Upsilon^2
+\left(\rho_\star+\frac{\beta_0}{3}-\frac{\beta_2}{M_\star^2}\right)
\Upsilon -\frac{\beta_1}{3M_\star^2} = 0\,.
\ee
This determines $\Upsilon$ as a function of $\rho_\star$.  The
analytic solutions are straightforward to derive but lengthy to
display. For generic values of parameters there is the further
complexity that in some cases one has to match different branches of
solutions in order to keep $\Upsilon$ real for all $\rho$, although a
real solution always exists. In order to keep the discussion
transparent we simply state some generic properties here and then
consider a few special cases in more detail.

First, at late times in an expanding universe, $\rho$ generically
approaches a constant value $\rho_{vac}$, the vacuum energy
density\footnote{Of course, we can always choose $\rho_{vac}=0$ and
  include the vacuum energy contribution entirely in the $\beta_0$
  parameter, or vice versa.}. Then from (\ref{quartic}) it is
obvious that at late times $\Upsilon$ approaches a constant value and
this holds for generic values of parameters. The Friedmann equations
(\ref{FgUpsilon}) and (\ref{FfUpsilon}) then imply that at late times
the universe always asymptotes towards a de Sitter or anti-de Sitter
geometry.

On the other hand, in the early time limit of $\rho_\star\rightarrow
\infty$ (in practice, $\rho_*>> \beta_n,\beta_n/M_*^2$, but not
necessarily including $\beta_0$) the linear term in (\ref{quartic})
dominates and forces a solution $\Upsilon\rightarrow 0$. The Friedmann
equation (\ref{FgUpsilon}) then implies an evolution for $H^2$
dominated by $\rho_*$ and a cosmological constant contribution
$\beta_0$, as in general relativity. Of course, even in this limit,
the full quartic equation has three more solutions (two being
complex), but these diverge in the limit $\rho_\star\rightarrow
\infty$, as can be inferred from the explicit form of the
solution. Thus, the solution that vanishes is physically preferable 
since we then get back ordinary general relativity very early in the
expansion history. While a priori this may not be necessary, it
guarantees that ordinary early universe considerations remain valid. 

As a caveat, note that while the discussion above is fomulated in
terms of $\Upsilon$ and holds in general, at the end, it is only the
behaviour of $H^2$ that concerns us. Later we will display an example
of an analytic solution where $\Upsilon$ diverges at early times while
the Friedmann equation for $H^2$ is completely equivalent to the
usual general relativistic equation.

To summarize, the equations admit generic solutions that start out
from a universe described by the ordinary general relativistic
Friedmann equation with the cosmological constant $\sim\beta_0m^2$,
and evolve toward a de Sitter (or AdS) universe, with a cosmological
constant depending on the parameters of the theory.

\subsection{Models with $\beta_3=0$}

In the following we consider primarily theories with
$\beta_3=0$. Although our main intent is to simplify the analysis of
the solutions we note that this choice is of interest as it
corresponds to neglecting the highest (cubic) order nonlinear
interactions in $\sqrt{g^{-1}f}$ in (\ref{action}). Also, in the
massive gravity limit, which corresponds to freezing the dynamics of
$f_{\mu\nu}$, only for this choice the authors in \cite{Koyama:2011yg}
find spherically symmetric cosmological solutions that exhibit the
Vainshtein mechanism in a manner consistent with
observations\footnote{Explicitly, \cite{Koyama:2011yg} find that for
  $\beta_3\neq0$,  the Vainshtein mechanism screens not only the
  scalar mode but also  the tensor modes and hence is ruled out.}.

For the choice $\beta_3=0$, (\ref{quartic}) reduces to a cubic
equation for $\Upsilon$ and one can integrate the $g_{\mu\nu}-$
Friedmann equation (\ref{FgUpsilon}) for $H$. Writing down the
explicit solution is not very illuminating, and we simply note that it
exists for generic $\beta_1$. This solution is fitted to data in the
following sections.

Here, however, to illustrate the general discussion of section
\ref{general}, we look for simpler analytic solutions. Note that the
more restrictive choice $\beta_3=0$ and $\beta_4=3\beta_2M_\star^2$
converts (\ref{quartic}) into a quadratic equation (the other
possibility $\beta_3=0$, $\beta_1=0$ is discussed below). Now,
solutions exist only for $\beta_1\neq0$,
\be
\Upsilon = -\,\frac{3\rho_\star+\beta_0-3\beta_2M_\star^{-2}}{6\beta_1}
\pm\frac{\Psi(\rho_\star)}{6\beta_1} \,,
\ee
where for convenience we have defined
\be
\Psi = \sqrt{(3\rho_\star+\beta_0-3\beta_2M_\star^{-2})^2+
12\beta_1^2M_\star^{-2}} \,.
\ee
Choosing the solution with the positive sign ensures that $\Upsilon
\rightarrow 0$ as $\rho\rightarrow\infty$, recovering standard early
cosmology as discussed above. For this choice of sign, the Friedmann
equation (\ref{FgUpsilon}) becomes,
\be
H^2+\frac{k}{a^2} = C_1\frac{\rho}{3M_g^2} + m^2C_2 
- \frac{m^2}{12}\left(2C_1-3\right)\Psi(\rho_\star) 
+ \frac{m^2\beta_2}{2\beta_1^2}\rho_\star^2 
- \frac{m^2\beta_2}{6\beta_1^2}\rho_\star\Psi(\rho_\star)\,,
\ee
where $C_{1,2}$ are constants given by
\begin{align}
C_1 &=\frac{1}{6\beta_1^2}\left(3\beta_1^2
+2\beta_0\beta_2-6\frac{\beta_2^2}{M_\star^2}\right)\,, \nn\\
C_2 &= \frac{1}{18\beta_1^2}\left(3\beta_0\beta_1^2+\beta_0^2\beta_2
+15\frac{\beta_1^2\beta_2}{M_\star^2}-6\frac{\beta_0\beta_2^2}{M_\star^2}
+9\frac{\beta_2^3}{M_\star^4}\right)\,.
\end{align}
Although this appears to be a highly nontrivial modification of the
general relativistic Hubble parameter, in the high energy limit of
large $\rho$ we have, 
\be
H^2+\frac{k}{a^2} \sim \frac{\rho}{3M_g^2} + \frac{m^2\beta_0}{3}\,,
\ee
while as the energy density dilutes towards a constant value
$\rho_{vac}$,  
\be
H^2+\frac{k}{a^2}\sim C_1\frac{\rho_{vac}}{3M_g^2}+m^2C_3\,. 
\ee
where the constant $C_3$ can be read off from the above expressions. 
This explicitly demonstrates the general arguments of section
\ref{general}.  

\subsection{Models with $\beta_1=\beta_3=0$}

Another particularly interesting class of solutions arise when both
$\beta_1=\beta_3=0$, and $\beta_2<\beta_4/3M_\star^2$. As discussed in
section \ref{review}, this corresponds to the minimal bimetric
model where the interactions are of the lowest order simultaneously
for both $\sqrt{g^{-1}f}$ and $\sqrt{f^{-1}g}$. This keeps only the
quadratic order nonlinear interactions in both sectors such that the
action (\ref{action}) looks like a nonlinear action with a mass
potential.

For these values we find the solution
\be
\Upsilon^2 = \frac{3\rho_\star+\beta_0
-3\beta_2M_\star^{-2}}{\beta_4M_\star^{-2}-3\beta_2}\,,
\ee
and the Friedmann equation (\ref{FgUpsilon}) is given by
\be\label{Fgb10b30}
H^2+\frac{k}{a^2} 
= \frac{\beta_4}{\beta_4-3\beta_2M_\star^2}\frac{\rho}{3M_g^2}
+\frac{m^2(\beta_0\beta_4-9\beta_2^2)}{3(\beta_4-3\beta_2M_\star^2)}\,.
\ee
In general, we can also split the energy density into matter and
vacuum contributions, $\rho=\rho_m +\rho_{vac}$. This equation is
completely degenerate with the ordinary general relativistic equation
with a rescaled Planck mass and a shifted cosmological constant. Thus
it can be easily fitted to cosmological data. From the observational
perspective any discrepancy with data must then be looked for by
examining the corresponding solutions at smaller scales where the
homogeneous solutions are not appropriate, e.g.~cluster scales. To make
this point more explicit and also consider neighbouring models with
$\beta_1\neq0$ we proceed to fit these models against observational
data from supernovae, cosmic microwave background, and baryon acoustic 
oscillations in the upcoming sections.

Note that this solution corresponds to a case where even though
$\Upsilon$ diverges for large $\rho$, the equation for $H$ has the
general relativistic form. This case evades the general arguments of 
section \ref{general}.

\section{Parameterization of the solution}\label{sec_parameter}

The bimetric action (\ref{action}) has a number of degenerate
parameters. In this section we discuss the choice of parameters to
facilitate comparison to data.

\subsection{Parameterization used for model fitting}
\label{parameter}

In order to make the comparison with standard cosmology more
transparent, we define
\be
M^2\equiv\frac{m^2}{H_0^2}\,,\quad E^2\equiv\frac{H^2}{H_0^2}\, ,
\ee
where $H_0$ is the present day Hubble parameter. We further define the 
density parameters, 
\be
\Omega \equiv\frac{\rho}{3M_g^2H_0^2}\,,\quad
\Omega_k\equiv-\frac{k}{a_0^2H_0^2}\,,\quad
\Omega_{\Upsilon}\equiv M^2\left(\frac{\beta_3}{3}\Upsilon^3 
+ \beta_2\Upsilon^2 + \beta_1\Upsilon
+ \frac{\beta_0}{3}\right)\, ,
\label{normalized}
\ee
where as usual
\be
\Omega = \Omega_\gamma(1+z)^4 + \Omega_m(1+z)^3 +\Omega_\Lambda +\dots
\ee
is given in terms of the fluid components, respectively, for
radiation, matter and vacuum etc., at redshift $z=0$. Dividing
through by $H_0^2$, the Friedmann equation then assumes the form 
\be
E^2 \equiv \Omega+\Omega_k(1+z)^2+\Omega_{\Upsilon}\, ,
\ee
where $E^2=1$ at redshift $z=0$ by definition. 

In the action (\ref{action}), the parameter $\beta_0$ is a
cosmological constant for the metric $g_{\mu\nu}$ and hence degenerate 
with the vacuum energy $\rho_{vac}$. This can be used to set
\be
\beta_0 = -3\beta_1-3\beta_2-\beta_3 \, .
\ee
The reason for this choice is simply to have similar conventions to
those sometimes used in massive gravity (where this choice eliminates
the contribution of the mass potential to the cosmological constant
but only for backgrounds where $g=f$). This parameter choice
factorizes $\Omega_{\Upsilon}$ as
\be
\Omega_{\Upsilon} = M^2\frac{\Upsilon-1}{3}\left[\beta_3(\Upsilon^2 
+\Upsilon+1)+3\beta_2(\Upsilon+1)+3\beta_1\right]\, .
\ee
Note that $\Upsilon_0=1$ is a consistent normalization that can be
achieved on rescaling $f_{\mu\nu}$ by adjusting $M_f^2$. This fixes,   
$\Omega+\Omega_k=1$ at the present epoch. The Friedmann equation for
$f_{\mu\nu}$ (\ref{FfUpsilon}) then allows us to determine, e.g.,
$\beta_4$ as,
\be
\beta_4 = -(\beta_1+3\beta_3) 
-3\left(\beta_2-\frac{k}{a_0^2m^2}-\frac{M_\star^2}{M^2}\right)\,.
\ee

Finally, note that the parameters $\beta_1,\beta_2,\beta_3$ are
degenerate with the mass scale $m^2$. In particular, we can fix,
\be
\beta_1 = -1-2\beta_2-\beta_3\,,
\ee
to render the Fierz-Pauli mass of the massive fluctuation independent
of the $\beta$'s. Then for this choice, 
\be\label{FP_mass}
m_{\mathrm{FP}}^2 = \frac{M_*^2+1}{M_*^2}m^2
\ee
is the Fierz-Pauli mass of the massive fluctuation when expanding the
metrics around a common background for canonically normalized
fluctuations. For $M_f>>M_g$ the $f_{\mu\nu}$ sector decouples and
$m^2$ becomes the mass for the massive fluctuation of $\gmn$.

\subsection{Model specific considerations}\label{Q}

We consider a model with zero spatial curvature, $k=0$, and
$\beta_3=0$. With the above choices for $\beta_0\,,\beta_1\,,\beta_4$,
these can now be parameterized as,
\bea
\beta_0 &=& 3+3\beta_2 = 6-3\alpha\nn\\
\beta_1 &=& -1-2\beta_2 = -3+2\alpha\nn\\
\beta_4 &=& 1-\beta_2 + 3\frac{M_*^2}{M^2} = \alpha + 3\frac{M_*^2}{M^2}
\eea
where we have defined
\be
\alpha\equiv 1-\beta_2\, .
\ee
Note that this is the $\bar\alpha$ parameter of \cite{Hassan:2011vm},
discussed further in Appendix \ref{app_parameters}. We can now write
the two Friedmann equations (\ref{FgUpsilon}) and (\ref{FfUpsilon}) in
terms of $\alpha$ as
\bea\label{eq:FE1new}
E^2=\Omega+(2-\alpha)M^2+(2\alpha-3)M^2\Upsilon+(1-\alpha)M^2\Upsilon^2\, ,
\eea
and 
\bea\label{eq:FE2new}
\left[\frac{M_*^2}{M^2}+\frac{\alpha}{3}\right]\Upsilon^3+
\left[1-\alpha-\frac{M_*^2}{M^2}E^2\right]\Upsilon
+\frac{2\alpha}{3}-1=0\, .
\eea
Since our goal is to obtain an analytical expression for $E(z)$, we first
substitute the expression for $E^2$ given in (\ref{eq:FE1new})
into (\ref{eq:FE2new}) to obtain
\begin{align}
&\left[\frac{\alpha}{3}+\frac{M_*^2}{M^2}+(\alpha-1)M_*^2\right]\Upsilon^3+
  (3-2\alpha)M_*^2\Upsilon^2
\nn\\
&\hspace{3.0cm} +\left[1-\alpha-\Omega\frac{M_*^2}{M^2}
+(\alpha-2)M_*^2\right]\Upsilon+\frac{2\alpha}{3}-1=0\, .
\label{eq:yba}
\end{align}
We then solve this cubic equation for $\Upsilon$ (being careful to
pick the solution corresponding to $\Upsilon_0=1$) and put this back
into the Friedmann equation (\ref{eq:FE1new}). The resulting solution
is fitted to data in the next sections.

In the case that $\alpha=3/2$ (corresponding to $\beta_1=0$ discussed
earlier), we obtain a particularly simple form for the expansion
history, 
\bea
E^2=\Omega\left[1-\Omega^{\mathrm{eff}}_{\Lambda}\right]+
  \Omega^{\mathrm{eff}}_\Lambda\,.
\label{eq:a15}
\eea
where, 
\be\label{Oeff}
\Omega^{\mathrm{eff}}_{\Lambda}=\frac{M^2M_*^2}{M^2+M_*^2(2+M^2)}\, .
\ee
appears as an effective cosmological constant and also contributes to
an effective Planck mass. To compare with data, we do not include any
vacuum contribution in $\Omega$ which, at late times, is then entirely
given by $\rho_m$. The actual outcome of this evolution equation
depends on the relation between the scale $M_g$ and the physical
Planck mass $M_P$ (or equivalently, the Newton constant $G_N$). In
general this will be of the form,
\be
M_P=Q\,M_g\,,
\ee
where $Q$ is given by the parameters of the theory. To explicitly
determine $Q$, we need localized bimetric solutions that are not well
understood at present, so we treat it as a parameter \footnote{It is
  easy to determine $Q$ in the linearized theory which will also
  exhibit a milder verion of vDVZ discontinuity. However, the real $Q$
  for non-linear solution will be different due to the Vainshtein
  effect.}. Then, in view of (\ref{normalized}) $\Omega$ is related to
the physical density parameter by 
\be
\Omega = \Omega_{\mathrm{phys}}\, Q^2
\ee
and the evolution equation becomes,
\be
E^2=\Omega_{\mathrm{phys}}\, Q^2 (1-\Omega^{\mathrm{eff}}_\Lambda )
+\Omega^{\mathrm{eff}}_\Lambda \,. 
\ee
Comparison to data, gives $\Omega^{\mathrm{eff}}_\Lambda =0.7$ so that
one always has $\Omega_{\mathrm{phys},0}\,Q^2=1$. The limiting values of $Q$
are $Q=1$ ($M_g=M_P$) and $Q^2 (1-\Omega^{\mathrm{eff}}_\Lambda )=1$
which is equivalent to the concordance model. 

As some limits, let us consider,
\bea\label{M_limit}
	M\rightarrow\infty\quad\Rightarrow\quad
  \Omega^{\mathrm{eff}}_{\Lambda}\rightarrow\frac{M_*^2}{1+M_*^2}\, ,
\eea
and
\bea\label{Ms_limit}
	M_*\rightarrow\infty\quad\Rightarrow\quad
  \Omega^{\mathrm{eff}}_{\Lambda}\rightarrow\frac{M^2}{2+M^2}\, .
\eea
This means that we always have a cosmological constant universe,
even as either $M$ or $M_*$ are extremely large, as long as the other
mass ratio is of the appropriate size. As an example, let us set
$Q=1$. Then the first of the above limits, eq.~(\ref{M_limit}),
corresponds to 
\be
M>>1\quad\Leftrightarrow\quad m>>H_0\, .
\ee
This, as can be seen from eq.~(\ref{FP_mass}), implies a very large mass for the spin-2 field measured in Hubble units. As is straightforward to
verify, given that $m/H_0>>1$ we need $M_f\sim 1.5M_g$ in order for
$\Omega^{\mathrm{eff}}_{\Lambda}$ to mimic $\ola\sim0.7$. It is indeed
interesting that this bimetric model can mimic the concordance model even for 
a very large mass for the massive spin-2 mode. It is not clear, however, how
this will effect smaller scale physics.
 
The second limit, eq.~(\ref{Ms_limit}), corresponds to
\be
	M_*>>1\quad\Leftrightarrow\quad M_f>>M_g\, .
\ee
This will effectively decouple the $f_{\mu\nu}$ field, making it a
free field determined by the vacuum Einstein equations. This in turn
makes the fluctuations of $\gmn$ massive with mass $m$, which again can
be seen from (\ref{FP_mass}). Now, in order for
$\Omega^{\mathrm{eff}}_{\Lambda}$ to mimic $\ola\sim0.7$, we need $m\sim2.2 H_0$.

We also note that the limit
\be
	M<<1\quad\Leftrightarrow\quad m<<H_0\, ,
\ee
can be seen from eq.~(\ref{Oeff}) to imply $\Omega^{\mathrm{eff}}_{\Lambda}\sim0$,
so that the limit of vanishing mass for the massive spin-2 field (for
non-zero $M_f$) gives no cosmological contribution.

\section{Data}\label{data}

In this study, we limit ourselves to purely geometrical tests of the
expansion history of the universe. That is, tests only involving
cosmological distances. We defer possible constraints involving
smaller scale gravity and structure formation to upcoming work. 

\subsection{Type Ia supernova data}\label{sec:sndata}
As being standardizable candles and thus effective distance
indicators, Type Ia supernovae (SNe~Ia) are one of the most direct
probes we have of the expansion history of the universe.
In this paper, we use the Union2
\cite{Amanullah:2010vv} compilation of SNe~Ia. 
This data set contains SNe~Ia from, e.g., the Supernova
Legacy Survey, ESSENCE survey and HST observations. After selection
cuts, the data set amounts to 557 SNe~Ia, spanning a redshift range of
$0\lesssim z \lesssim 1.4$, analyzed in a homogeneous fashion using
the spectral-template based fit method SALT2.

\subsection{Cosmic Microwave Background and Baryon Acoustic Oscillations}

The position of the first Cosmic Microwave Background (CMB)
power-spectrum peak, representing the angular scale of the sound
horizon at the era of recombination, is given by
\bea 
  \ell_A = \pi \frac{d_{\rm A}(z_*)}{r_{\rm s}(z_*)}\, ,
\eea
where $d_{\rm A}(z_*)$ is the comoving angular-diameter distance to
recombination while the comoving sound horizon at photon decoupling,
$r_{\rm s}$, is given by
\bea 
  r_{\rm s} = \int_{z_*}^\infty \frac{c_s}{H(z)}dz\, ,
  \label{eq:rs}
\eea 
which depends upon the speed of sound before recombination, $c_s$.
Here we use CMB measurements from the seven-year Wilkinson
Microwave Anisotropy Probe (WMAP) observations
\cite{Komatsu:2010fb}, in this case the WMAP7.2 results 
reported at lambda.gsfc.nasa.gov, adopting the value
$\ell_A=302.56\pm0.78$. We further assume $z_*=1091.12$ exactly
(variations within the uncertainties about this value do not give
significant differences to the results).

Baryon Acoustic Oscillations (BAO) observations are often compared to
theoretical models using measurements of the ratio of the sound
horizon scale at the drag epoch, $r_s(z_d)$, to the dilation scale,
$D_V(z)$.  The drag epoch, $z_d\approx1020$, is the epoch at which the
acoustic oscillations are frozen in. A more model-independent
constraint can be achieved by multiplying the BAO measurement of
$r_s(z_d)/D_V(z)$ with the CMB measurement $\ell_A=\pi
d_A(z_*)/r_s(z_*)$, thus cancelling some of the dependence on the
physical size of the sound horizon scale \cite{Sollerman:2009yu}. In
doing this, we are effectively only left with the assumption that
the observed inhomogeneities in the large scale distribution of
galaxies and in the CMB temperature reflects the same (redshifted)
physical scale.

In \cite{Percival:2009xn}, measurements of the ratio $r_s(z_d)/D_V(z)$
at two redshifts, $z=0.2$ and $z=0.35$, are reported as
$r_s(z_d)/D_V(0.2)=0.1905\pm0.0061$ and
$r_s(z_d)/D_V(0.35)=0.1097\pm0.0036$.  Before matching to cosmological
models, we need to implement a correction for the difference between
the sound horizon at the end of the drag epoch, $z_d\approx1020$, and
the sound horizon at last-scattering, $z_*\approx1091$, the first
being relevant for the BAO and the second for the CMB. Here, we use
$r_s(z_d)/r_s(z_*)=1.0451\pm0.0158$, again using WMAP7.2 results. A
possible caveat is that this ratio was calculated using standard
cosmology for the evolution between the two redshifts. However, we
expect this to be a good approximation since the redshift difference
is relatively small, and the sound horizon at decoupling and drag is
mostly governed by the fractional difference between the number of
photons and baryons. Combining this with $\ell_A$ gives the numbers we
employ in our cosmology fits,
\bea
\frac{d_A(z_*)}{D_V(0.2)} &=&  17.55\pm0.62\, ,\label{eq:dADV_drag}\\
\frac{d_A(z_*)}{D_V(0.35)} &=& 10.11\pm0.34\, . \nonumber
\eea
We take into account the correlation between these measurements using
a correlation coefficient of 0.337 calculated in
\cite{Percival:2009xn}.

\section{Results of observational tests}\label{results}

\subsection{$\alpha=3/2$ ($\beta_1=0$)}
\begin{figure}
\begin{center}
\includegraphics[angle=0,width=.49\textwidth]{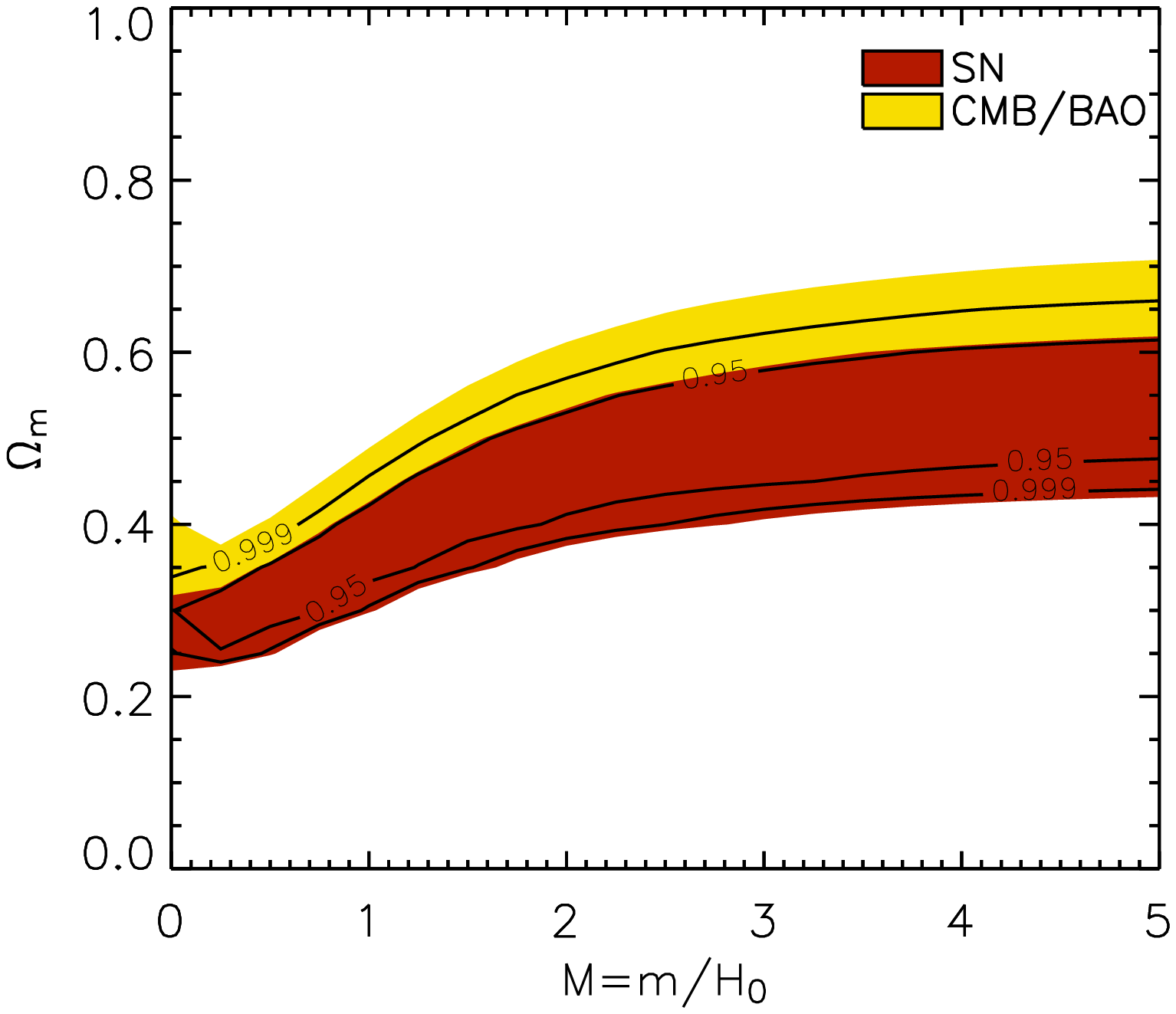}
\includegraphics[angle=0,width=.49\textwidth]{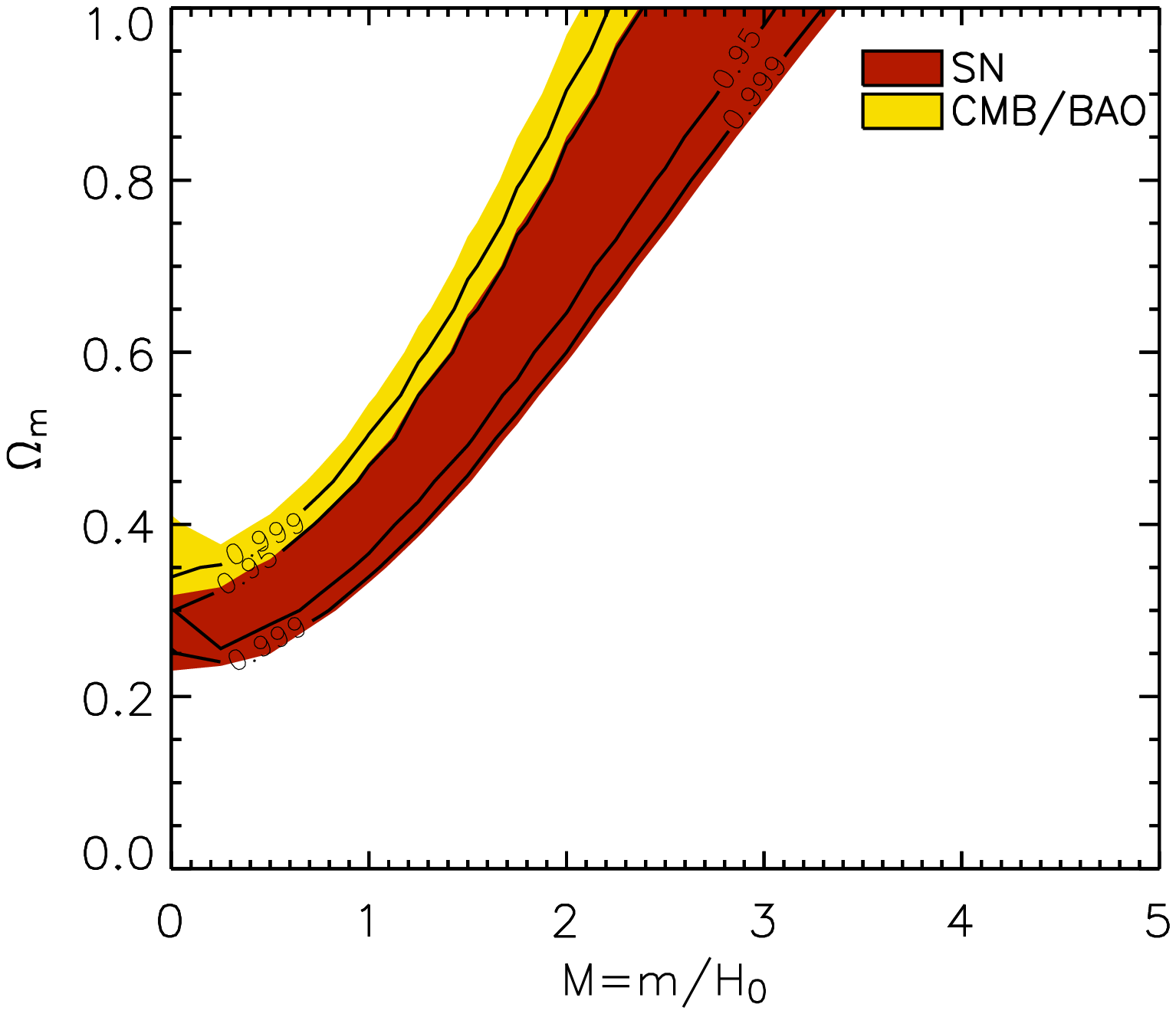}
\caption{\label{fig:a15om} Cosmological constraints for $\alpha = 3/2$ 
using supernova distances (SN) and the ratio of the observed scales of
the baryon acoustic oscillations as imprinted in the cosmic microwave
background and the large scale galaxy distribution (CMB/BAO). In the
left panel, a value of $M_*=1$ is assumed, in the right panel,
$M_*=3$.}
\end{center}
\end{figure}

If we fix the value of $M_*=M_f/M_g$ and allow for a cosmological
constant, we can (assuming a flat universe) fit for the spin-2 mass
parameter $m$ 
and the matter density $\om$. We expect to get a good fit for $M=0$
and $\om\sim 0.3$ since this corresponds to the concordance cosmology
with $\ola=0.7$. If we are able to get a good fit also when
$\om\rightarrow 1$, that is with no cosmological constant, depends on
the value of $M_*$.  From Figure~\ref{fig:a15om}, we see that in the
case of $M_*=1$ (left panel), this is not possible, while in the
case of $M_*=3$ (right panel), it is, as expected from
Eq.~(\ref{eq:a15}).  In this and all figures hereafter, shaded
contours shows constraints for SN and CMB/BAO data, respectively,
corresponding to $95\,\%$ confidence interval for two
parameters. Combined constraints are shown with solid lines
corresponding to $95\,\%$ and $99.9\,\%$ confidence intervals for two
parameters.
%----------------------------------------------------------------------
%----------------------------------------------------------------------

\begin{figure}
\begin{center}
\includegraphics[angle=0,width=.49\textwidth]{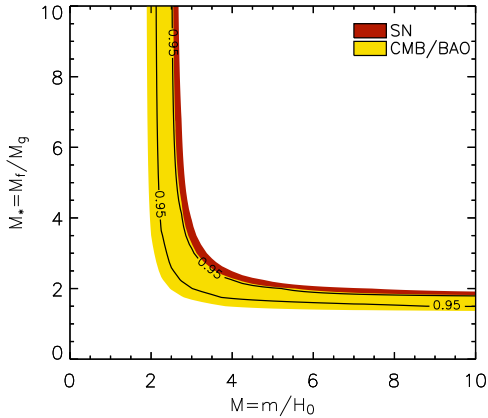}
\caption{\label{fig:a15} Cosmological constraints in the
  $[M,M_*]$-plane assuming $\alpha = 3/2$ and $\om=1$.}
\end{center}
\end{figure}

In what follows, we will set $\om=1$ and $\ola=0$ in order to
investigate whether bimetric gravity models can explain the apparent
accelerated expansion seen in cosmological geometrical data.  Fitting
both $M_*$ and $M$, we expect to be able to obtain good fits to the
data as long as the combination
$\Omega^{\mathrm{eff}}_{\Lambda}=M^2M_*^2/(M^2+M_*^2(2+M^2))$ is close
enough to the concordance value for the cosmological constant. As
discussed previously, this can always be achieved if either one of the
values is large enough, see Figure~\ref{fig:a15}.
%----------------------------------------------------------------------
%----------------------------------------------------------------------

\subsection{General $\alpha$ ($\beta_1\neq0$)}
\begin{figure}
\begin{center}
\includegraphics[angle=0,width=.49\textwidth]{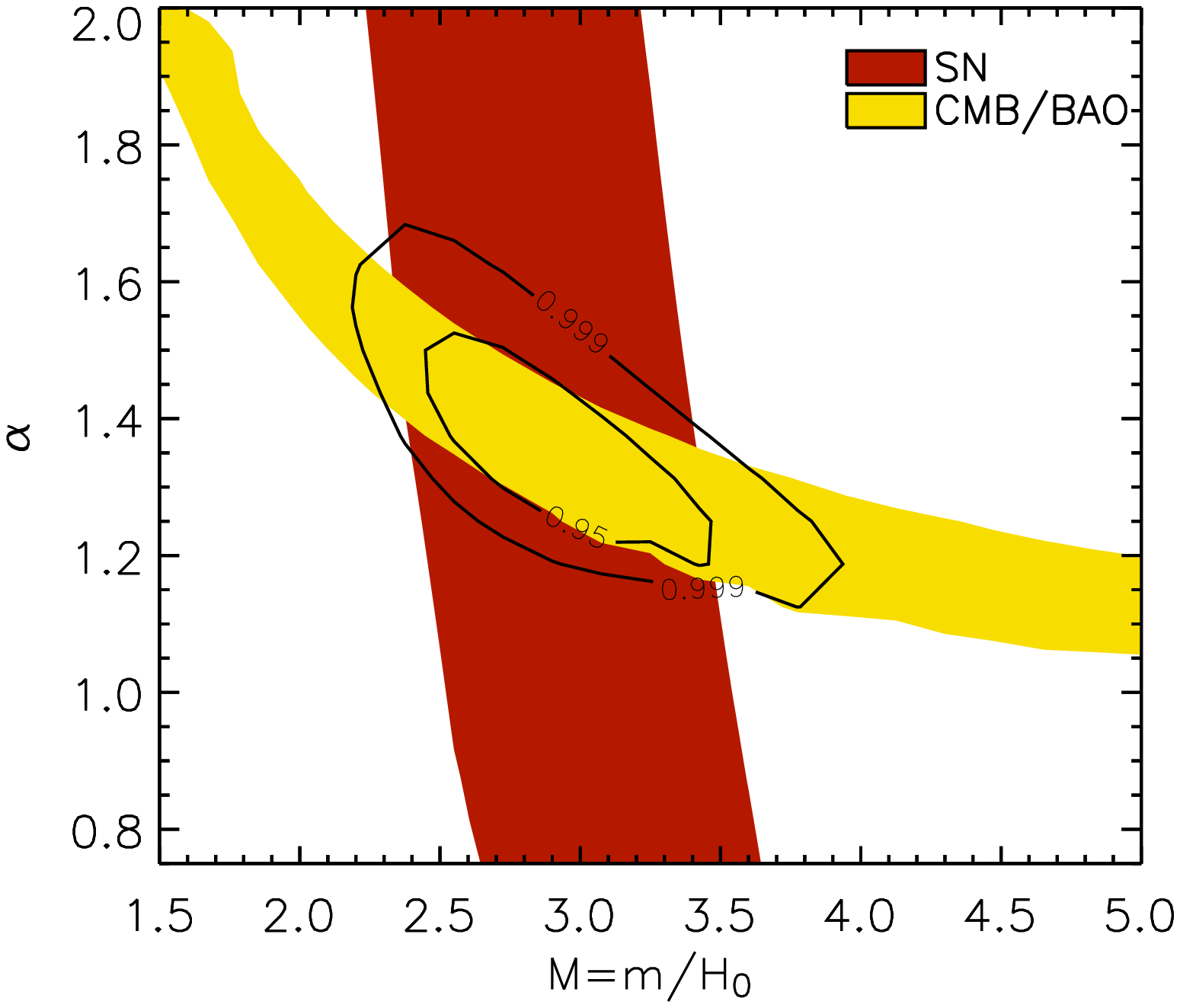}
\includegraphics[angle=0,width=.49\textwidth]{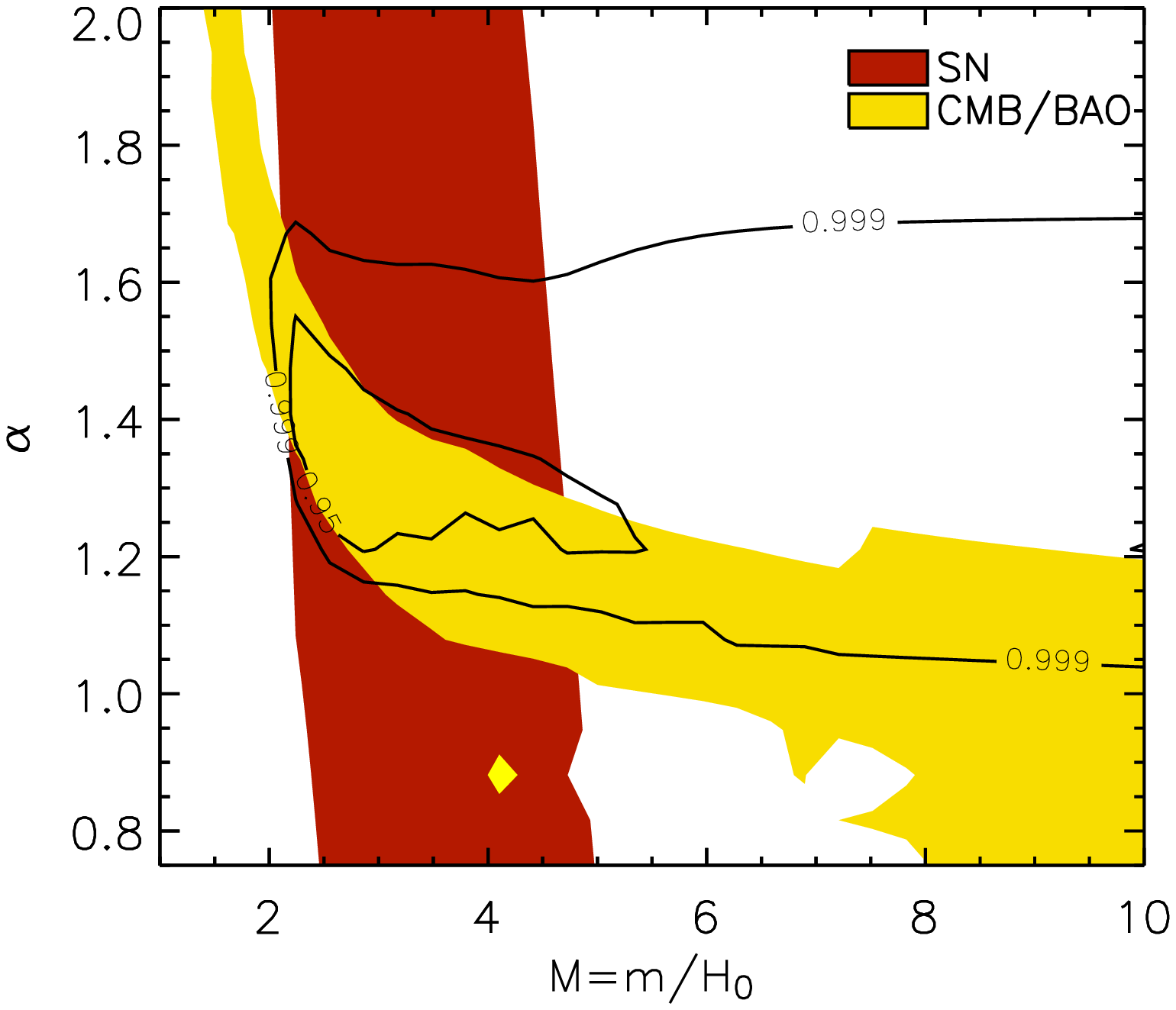}
\caption{\label{fig:aMMscr} {\it Left panel:} Cosmological constraints
  in the $[\alpha,M]$-plane assuming 
$\om=1$ and $M_*=3$. {\it Right panel:} Constraints in the
$[\alpha,M]$-plane for $\om=1$ after marginalizing over $M_*$ with a
flat prior in the interval $[0,10]$.}
\end{center}
\end{figure}

We now proceed to fit also the value of $\alpha$ (assuming
$\ola=0$). For $M_*=3$, we obtain the result depicted in the left
panel of Figure~\ref{fig:aMMscr} showing a good fit to the data for
$\alpha\sim 1.4$ and $M\sim 3.0$.

Even more generally, we want to fit $\alpha$, $M$ and $M_*$
simultaneously. This requires care when projecting results on two
dimensional surfaces. Since we are able to obtain good fits to the
data for both $M\rightarrow\infty$ and $M_*\rightarrow\infty$, we will
not be able to cover the entire non-negligible probability function in
our grid of tested parameter values which in principle is required to
perform a proper marginalization. This will mostly affect the
projected constraints for $M$ and $M_*$ where generally a larger
parameter space in one of the parameters will give more weight to the
likelihood function of the other parameter at lower values (see
Figure~\ref{fig:a15}). In the following, we present results for $M^{\rm
max}=M_*^{\rm max}=10$, the equivalent of putting a flat prior on the
values of $M$ and $M_*$ to be in the interval $[0,10]$. Results in the
$[\alpha,M]$-plane after marginalizing over $M_*$ are shown in the
right panel of Figure~\ref{fig:aMMscr}. 
%----------------------------------------------------------------------
%----------------------------------------------------------------------
Comparing to the left panel, we can see that allowing $M_*$ to vary
slightly shifts and widens the allowed values of $\alpha$ and $M$ as
compared to the case of a fixed value of $M_*$. Marginalizing down to one
parameter surfaces (assuming flat prior probablilities), our data
constrains the parameter values of the bimetric gravity model to be
(at $95\,\%$ confidence level for one parameter)
\be
1.1\lesssim\alpha\lesssim 1.5\,,\quad2\lesssim M\lesssim 3.5\,,\quad
1.5\lesssim M_*\lesssim 3\, .
\ee

\subsection{Interpretation of results}

As expected from the form of the cosmological solutions, the bimetric
gravity model is perfectly capable of matching geometric cosmological
data. For the case of $\alpha=3/2$ ($\beta_1=0$), this was obviously
going to be possible for some values, but we find that data clearly
favours the regions $M\sim2.5\,,M_*\gtrsim1.5$ or
$M\gtrsim2.5\,,M_*\sim1.5$ (assuming $M_P=M_g$ for simplicity, the
more general case having been discussed in section \ref{Q}). Allowing
for an arbitrary $\alpha$ is equivalent to switching on the $\beta_1$
interactions, with the only restriction that we focus here on
bigravity models for which $\beta_1=-1-2\beta_2$, which include
massive gravity limits of bigravity. In this case, we find that data
favours a rather narrow region of parameter values. In particular,
large deviations from $\alpha=3/2$ as well as $m=3H_0$ and $M_f=2M_g$
are being disfavoured by data.

\section{Conclusions}\label{discussion}

We have investigated cosmological solutions in the unique classically
consistent theory of bimetric gravity. Under the assumption that both
metrics respect the symmetries of spatial isotropy and homogeneity, we
derived the most general cosmological evolution equations for this
theory. The generic solution was demonstrated to always allow for a
cosmic evolution starting out from an ordinary FLRW matter dominated
universe while evolving towards a de Sitter (or AdS) geometry at late
times in the expansion history. We explored further a particular class
of solutions, corresponding to neglecting cubic nonlinear interactions
in the potential for the two metrics. A subclass of these solutions,
the minimal model, was shown to be completely degenerate with the
evolution of a universe described by the usual cosmic concordance
model in general relativity. Using recent data from SNe~Ia, CMB and
BAO we demonstrated that data favoured regions in parameter space
close to this minimal model. Interestingly, assuming $M_P\sim M_g$, we
found that even for a very large mass for the massive spin-2 mode can
this theory match observations as good as general relativity, although
under the assumption of order unity values for the relative couplings,
data favoured a mass of the order of the Hubble scale $m\sim H_0$.

Since the bimetric gravity theory is a highly nontrivial but
consistent modification of general relativity, it is important to
explore its consequences further. In particular, exploring whether
more general classes of solutions than the subclass studied in this
paper are able to match observations. Since we have demonstrated the
existence of bimetric solutions that match general relativity on
cosmic scales it is important to study perturbations on these
solutions and also obtain the corresponding small scale solutions in
order to see what this theory predicts for observations on e.g.
cluster and galaxy scale. Another important issue is the
determination of the observed Planck scale in terms of the parameters
of the theory.

\acknowledgments

We would like to thank Bo Sundborg, Rachel Rosen, Stefan Sj\"{o}rs,
for useful discussions and comments. We are indebted to M. Crisostomi
for pointing out an errounous statement in our first draft. EM
acknowledges support for this study by the Swedish Research Council.

\vspace{.5cm}

\noindent {\bf Note Added:} When the writing of this paper was being
finalized, a paper with similar intent appeared \cite{Volkov:2011an}
with some overlap with section 3 of the present paper. After this
paper was submitted to the arXiv, \cite{Comelli:2011zm} appeared which also
considers cosmological implications of bimetric gravity
\vspace{.3cm}
\appendix

\section{Derivation of the equations of motion}\label{app_eom}

In this appendix we give some details on the derivation of the
equations of motion, equations (\ref{Fg}), (\ref{Accg}), (\ref{Ff}),
and (\ref{Accf}). We insert our ansatz for the two metrics
(\ref{ansatz}), into the $g$- and $f$-equations of motion
(\ref{g_eom}) and (\ref{f_eom}), respectively. With our ansatz the
$00$-component of the $\gmn$ equation of motion (\ref{g_eom}) becomes  
\bea
-3\frac{\dot{a}^2}{a^2}-3\frac{k}{a^2}+m^2\left(\beta_0+3\beta_1
\frac{Y}{a}+3\beta_2\frac{Y^2}{a^2}+\beta_3\frac{Y^3}{a^3}\right)
=\frac{{T^0}_0}{M_g^2}\, ,
\eea
and the $11$-component reads
\bea\label{accgg}
-2\frac{\ddot{a}}{a}&-&\frac{\dot{a}^2}{a^2}-\frac{k}{a^2}+m^2\left(
\beta_0+\left[2\frac{Y}{a}+X\right]\beta_1+\left[\frac{Y^2}{a^2}
  +2\frac{XY}{a}\right]\beta_2+\frac{XY^2}{a^2} \beta_3 
\right)
=\frac{{T^1}_1}{M_g^2}\,. \nonumber \\
\eea
Now we insert the solution of the Bianchi constraint,
\be
	X = \frac{\dot Y}{\dot a} = \frac{\td Y}{\td a},
\ee
to rewrite (\ref{accgg}) as
\bea
-2\frac{\ddot{a}}{a}&-&\frac{\dot{a}^2}{a^2}-\frac{k}{a^2}+m^2\left(\beta_0
	+\left[2\frac{Y}{a}+\frac{\dot{Y}}{\dot{a}}\right]\beta_1+
        \left[\frac{Y^2}{a^2}+2\frac{\dot{Y}Y}{\dot{a}a}\right]
        \beta_2+\frac{\dot{Y}Y^2}{\dot{a}a^2}\beta_3\right)   
=\frac{{T^1}_1}{M_g^2} .\nonumber \\
\eea
In order to compute the contribution from the kinetic terms to the
$f_{\mu\nu}$ equations of motion~(\ref{f_eom}), we need the expression
for the Ricci tensor $R_{\mu\nu}(f)$ in terms of the two scale
factors. The non-vanishing Christoffel symbols for $f_{\mu\nu}$ are
\bea
\Gamma^0_{00}=\frac{\dot{X}}{X}\,,\quad
\Gamma^0_{11}=\frac{\dot{Y}Y}{X^2(1-kr^2)}\,,\quad
\Gamma^0_{22}=\frac{\dot{Y}Yr^2}{X^2}\,,\quad
\Gamma^0_{33}=\frac{\dot{Y}Yr^2\sin^2\theta}{X^2}\,,\nonumber\\
\Gamma^i_{0i}=\frac{\dot{Y}}{Y}\,,\quad\quad
\Gamma^1_{11}=\frac{kr}{1-kr^2}\,,\quad
\Gamma^1_{22}=-r(1-kr^2)\,,\quad
\Gamma^2_{12}=\Gamma^3_{13}=\frac{1}{r}\,,\quad
\nonumber\\
\Gamma^1_{33}=-r\sin^2\theta(1-kr^2)\,,\quad
\Gamma^2_{33}=-\sin\theta\cos\theta\,,\quad
\Gamma^3_{23}=\cot\theta\,.\quad
\eea
From these we compute the non-vanishing components of the Ricci tensor,
\bea
R_{00}&=&\frac{3}{Y^2}\left(-Y\ddot{Y}+\frac{Y\dot{Y}\dot{X}}{X}\right)\,,\nonumber\\
R_{11}&=&\frac{1}{X^2(1-kr^2)}\left(Y\ddot{Y}-\frac{Y\dot{Y}\dot{X}}{X} +2\dot{Y}^2 +2kX^2\right)\,,\nonumber\\
R_{22}&=&\frac{r^2}{X^2}\left(Y\ddot{Y}-\frac{Y\dot{Y}\dot{X}}{X} +2\dot{Y}^2 +2kX^2\right)\,,\nonumber\\
R_{33}&=&\frac{r^2\sin^2\theta}{X^2}\left(Y\ddot{Y}-\frac{Y\dot{Y}\dot{X}}{X} +2\dot{Y}^2 +2kX^2\right)\,.
\eea
Thus, the curvature scalar is given by
\bea
R=\frac{6}{X^2}\left(\frac{\ddot{Y}}{Y}-\frac{\dot{Y}\dot{X}}{YX}+\frac{\dot{Y}^2}{Y^2} +\frac{kX^2}{Y^2}\right).
\eea
The Einstein tensor then has the nonvanishing components
\bea
R_{00}-\frac{1}{2}f_{00}R&=&3\frac{\dot{Y}^2}{Y^2}+3\frac{kX^2}{Y^2}\,,\nonumber\\
R_{11}-\frac{1}{2}f_{11}R&=&-\frac{1}{X^2(1-kr^2)}\left(2\ddot{Y}-2\frac{Y\dot{Y}\dot{X}}{X}+\dot{Y}^2+kX^2\right)\,,\nonumber\\
R_{22}-\frac{1}{2}f_{22}R&=&-\frac{r^2}{X^2}\left(2\ddot{Y}-2\frac{Y\dot{Y}\dot{X}}{X}+\dot{Y}^2+kX^2\right)\,,\nonumber\\
R_{33}-\frac{1}{2}f_{33}R&=&-\frac{r^2\sin^2\theta}{X^2}\left(2\ddot{Y}-2\frac{Y\dot{Y}\dot{X}}{X}+\dot{Y}^2+kX^2\right)\,.
\eea
For the equations of motion, we raise one index with $f^{\mu\nu}$ and consider the $00$- as well as the $11$-component . The $00$-component of the equations of motion is then obtained as
\bea\label{eomzerof}
-3\frac{\dot{Y}^2}{X^2Y^2}-3\frac{k}{Y^2}+\frac{m^2}{M_*^2}\left[\frac{a^3}{Y^3}\beta_1+3\frac{a^2}{Y^2}\beta_2+3\frac{a}{Y}\beta_3+\beta_4\right]=0,
\eea
whereas the $11$-component reads
\bea\label{treomf}
0&=&-\frac{1}{X^2}\left(2\frac{\ddot{Y}}{Y}-2\frac{\dot{Y}\dot{X}}{YX}+\frac{\dot{Y}^2}{Y^2}\right)-\frac{k}{Y^2}\nonumber\\
&~&+\frac{m^2}{M_*^2}\left[\frac{a^2}{XY^2}\beta_1+\left(\frac{a^2}{Y^2}+\frac{2a}{XY}\right)\beta_2+\left(2\frac{a}{Y}+\frac{1}{X}\right)\beta_3+\beta_4\right].\nonumber\\
\eea
Inserting  $X=\dot{Y}/\dot{a}$ into (\ref{eomzerof}) and (\ref{treomf}) we arrive at
\bea
0&=&-3\frac{\dot{a}^2}{Y^2}-3\frac{k}{Y^2}+\frac{m^2}{M_*^2}\left[\frac{a^3}{Y^3}\beta_1
	+3\frac{a^2}{Y^2}\beta_2+3\frac{a}{Y}\beta_3+\beta_4\right]\,, \label{eins}\\
0&=&-2\frac{\dot{a}\ddot{a}}{Y\dot{Y}}-\frac{\dot{a}^2}{Y^2}-\frac{k}{Y^2}     +\frac{m^2}{M_*^2}\left[\frac{a^2\dot{a}}{\dot{Y}Y^2}\beta_1+\left(\frac{a^2}{Y^2}+\frac{2a\dot{a}}{\dot{Y}Y}\right)\beta_2+\left(2\frac{a}{Y}+\frac{\dot{a}}{\dot{Y}}\right)\beta_3+\beta_4\right]\,.\nonumber  \\\label{zwei} 
\eea
We now observe that acting with $(3+(Y/\dot Y)\p_t)/3$ on (\ref{eins})
gives (\ref{zwei}). Thus, the two equations are equivalent. 

\section{Comments related to massive gravity}\label{app_massgrav}

In order to fascilitate a comparison with other works on both bimetric
and massive gravity, using the consistent interaction term of the
action (\ref{action}), we note here the relation between the different
parameters of the action that has been established in the recent
literature on massive gravity, in particular
\cite{deRham:2010kj,Hassan:2011vm}. We also study the nature of the
Bianchi constraint and clarify some points related to cosmological
solutions recently obtained in the massive gravity literature.

\subsection{Parameters of massive gravity}\label{app_parameters}
In massive gravity, the metric that does not couple to matter is
regarded as a fixed background, usually taken to be a flat reference
metric (although this is not necessary). This implies removing the
kinetic strength $M_f$. One must also fix
$\beta_0=-3\beta_1-3\beta_2-\beta_3$ in order to cancel terms linear
in perturbations around this background. In order for $m^2$ to
correspond to the mass of the massive spin-2 mode one has to fix also
$\beta_1=-1-2\beta_2-\beta_3$. This effectively eliminates three
parameters. Since $\beta_4$ does not enter the equations of motion for
$g_{\mu\nu}$ one can consistently eliminate this parameter as well.
Thus, out of the five $\beta_n$ parameters of the general bimetric
theory only two are important for massive gravity. One is left with
four free parameters (including the Planck mass and the spin-2 mass).
The remaining two parameters in the interaction can be chosen
arbitrarily and (at least) two conventions have occured in the
literature. First, we note the relations between $\beta_n$ and the
$\bar\alpha_n$ used in \cite{Hassan:2011vm} 
\bea
	\beta_0 &=& 6-4\bar\alpha_3+\bar\alpha_4\,,\nn\\
	\beta_1 &=& -3+3\bar\alpha_3-\bar\alpha_4\,,\nn\\
	\beta_2 &=& 1-2\bar\alpha_3+\bar\alpha_4\,,\nn\\
	\beta_3 &=& \bar\alpha_3-\bar\alpha_4\,.
\eea
The relation between these and the parameters of \cite{deRham:2010kj} is
\bea
	\bar\alpha_3 &=& -3\alpha_3 = 6 c_3\,,\nn\\
	\bar\alpha_4 &=& 12\alpha_4 = -48 d_5\,.
\eea

\subsection{The Bianchi constraint}

We recall the Bianchi constraint (\ref{bianchi_constraint}) 
\bea
\label{app_bianchi_constraint}
\frac{3m^2}{a}\left[\beta_1 +2\frac{Y}{a}\beta_2+\frac{Y^2}{a^2}
\beta_3\right]\left(\dot{Y}-\dot{a}X\right)=0.
\eea
Clearly we can enforce this by looking for solutions for $Y$ which force the left bracket to vanish. This will however only result in the ordinary general relativistic equations for $\gmn$ with the addition of a cosmological constant proportional to $m^2$ as is evident from (\ref{Fg}), independent of any dynamics for $f_{\mu\nu}$. As such they represent a particular class of screening solutions with special values for the parameters of the action. 

The reason these values are special can be quantified further. If we consider perturbations around solutions where the metrics are proportional, i.e.
\be
	g = \bar g +\delta g\,,\quad f = C\bar g + \delta f\,,
\ee
we have that
\be
	g^{-1}f \approx C + \bar g^{-1}\left(\delta f 
		- C\delta g\right) \equiv C + \delta M\,,
\ee
where $\bar g\delta M$ defines the massive fluctuations up to a constant of proportionality. Using this and expanding the interaction term in the equations of motion (\ref{g_eom}) to linear order we obtain (excluding linear contributions from the cosmological term)
\bea
	\sum_{n=0}^3(-1)^n\beta_n
         g_{\lambda(\mu}Y_{(n)\nu)}^\lambda(\sqrt{g^{-1}f}) &\approx&
		\left(\beta_1+2C\beta_2+C^2\beta_3\right)
		\left[\bar g\left(\mathrm{Tr}(\delta M)
		-\delta M\right)\right]_{(\mu\nu)}\,.
\eea
This is just the Fierz-Pauli mass contribution to the linearized equations of motion. Comparing this to the Bianchi constraint (\ref{app_bianchi_constraint}) we see that forcing the left bracket of the Bianchi constraint to vanish for proportional metrics is equivalent to choosing a constant of proportionality such that the Fierz-Pauli mass term for the fluctuations vanish. A similar conclusion holds also for the cosmological solutions when the spatial metrics are proportional but this requires more work to show than the simple example discussed here.

For our ansatz (\ref{ansatz}), the Bianchi constraint (\ref{app_bianchi_constraint}) allows to be less strict and only demand that the spatial part of the metrics are proportional, i.e. $Y(t)=C a(t)$. Then by forcing the left bracket of the Bianchi constraint to vanish we can find bimetric solutions that read
\be\label{Ypropa}
	Y^2(t) = C^2a^2(t)\,,\quad
	X^2(t) =\frac{3C^2M_*^2\dot a^2}{-3k M_*^2
	+ m^2\left(\beta_2+C(2\beta_3+C\beta_4)\right))a^2}\,,
\ee
where $X$ is determined from the $f_{\mu\nu}$ equations of motion (\ref{f_eom}) and the constant $C$ is determined from the Bianchi constraint (\ref{app_bianchi_constraint}) to be given by
\be
	C_\pm = -\frac{\beta_2\pm\sqrt{\beta_2^2-\beta_1\beta_3}}{\beta_3} \,.
\ee
Although the existence of these solutions might appear interesting, in the bimetric theory where we include dynamics for $f_{\mu\nu}$, they do not modify general relativity in any respect apart from the contribution of a cosmological source given explicitly by the addition
\be
	m^2\left[\beta_0+\frac{1}{\beta_3}\left(2\frac{\beta_2^3}{\beta_3}-3\beta_1\beta_2
	\pm2\frac{\beta_2^2}{\beta_3}\sqrt{\beta_2^2-\beta_1\beta_3}
	\mp2\beta_1\sqrt{\beta_2^2-\beta_1\beta_3}\right)\right]\,,
\ee
to the usual Friedmann and acceleration equations of general relativity without a cosmological constant. From our previous reasoning this can be understood as a consequence of using the parameters of the general theory to impose the vanishing of the massive fluctuations. Hence, the theory contains only massless spin-2 fluctuations and must be equivalent to ordinary general relativity. Since we keep all parameters arbitrary such solutions could be expected, and a very similar conclusion was reached when examining spherically symmetric solutions in the consistent bimetric theory of gravity in \cite{Comelli:2011wq}.

If we do not include a kinetic term for $f_{\mu\nu}$, as in the massive gravity setups, these solutions do allow for regular cosmological evolution for $\gmn$, but again only contribute with a constant source addition (c.f. (\ref{f_eom})). For such a scenario $X$ is not given as in (\ref{Ypropa}), since we have used the $f_{\mu\nu}$ equations of motion to derive that solution. Indeed, without the dynamical term for $f_{\mu\nu}$, $X$ can be an arbitrary function of time and in particular it is possible to choose $X$ such that $f_{\mu\nu}$ represents an open chart of the Minkowski metric, as was recently demonstrated in \cite{Gumrukcuoglu:2011ew}. From this perspective it is also clear why it is not possible to find solutions with positive or zero spatial curvature for a non-dynamical $f_{\mu\nu}$, there is simply no representation of Minkowski space as a homogeneous and isotropic space with nontrivial scale factor for these cases.

\subsection{$X=constant$}
More generally, we recall the true dynamical Bianchi constraint (\ref{bianchi_constraint_dyn}),
\be\label{app_bianchi_constraint_dyn}
	X = \frac{\dot Y}{\dot a} = \frac{\td Y}{\td a}\,,
\ee
obtained from the vanishing of the right bracket of (\ref{app_bianchi_constraint}). Note that this encodes also the $Y\propto a$ solutions obtainable from enforcing the vanishing of the left bracket, but in this case will enforce $f_{\mu\nu}\propto g_{\mu\nu}$. Thus, for arbitrary parameters of the theory (\ref{app_bianchi_constraint_dyn}) contains all non-trivial information about the nature of the constraint. This allow for the simple class of solutions
\be
	Y = C_1 a + C_2\quad\Rightarrow\quad\Upsilon = C_1 + \frac{C_2}{a}\,.
\ee
where $C_{1,2}$ are constants. The Friedmann equation for $\gmn$ (\ref{g_eom}) is then given by
\begin{align}
	&\frac{H^2}{m^2}+\frac{k}{a^2m^2} = \rho_\star
		+\frac{\beta_0}{3}+C_1\left(\beta_1+\beta_2C_1+\frac{\beta_3C_1^2}{3}\right)
		+\left(\beta_1+2\beta_2C_1+\beta_3C_1^2\right)\frac{C_2}{a}\nn\\
		&\hspace{+3.8 cm}+\left(\beta_2+\beta_3C_1\right)\frac{C_2^2}{a^2}
		+\frac{\beta_3}{3}\frac{C_2^3}{a^3}\,.
\end{align}
For $C_2=0$ this clearly only amounts to a cosmological constant contribution to the usual general relativistic equations. Moreover, if $C_1=1$ we see that the choice $\beta_0=-3\beta_1-3\beta_2-\beta_3$ completely eliminates this contribution (c.f. discussion in Appendix \ref{app_parameters}).

More interestingly, if $C_2\neq0$ the addition to the general relativistic Friedmann equation is precisely of the form to add extra fluid components of the conventional type. From the perspective of the usual formulation of massive gravity then, where there is no dynamical equation for $f_{\mu\nu}$, these solutions can be interesting since they can exist for an arbitrary source and give a contribution to all the fluid components except for radiation. Solutions of this type were recently discussed in that context in \cite{Chamseddine:2011bu}. The added degeneracy in parameter space is however a rather unpleasant feature.

In the context of bimetric theory that we are considering here, the equations of motion for $f_{\mu\nu}$ (\ref{FfUpsilon}) imply that these solutions will constrain the source, as is evident from the quartic equation (\ref{quartic}). In fact, they will tell us that the source will contain terms of the form
\be\label{Xconst}
	\frac{a}{C_1a+C_2}\,,\quad\frac{C_2}{C_1a+C_2}
	\,,\quad\frac{C_2}{a(C_1a+C_2)}\,,\quad\frac{C_2}{a^2(C_1a+C_2)}\,,
\ee
up to various multiplicative constants. For $C_2=0$, they are of the standard matter fluid type but then only the constant contribution can remain, as we have also remarked on earlier. For $C_1=0$ (such that $f_{\mu\nu}$ is flat) the Bianchi constraint (\ref{app_bianchi_constraint_dyn}) tell us that either $X=0$ or $a$ is a constant and that there can be no cosmological evolution and hence this case is not interesting for our purposes. This conclusion was reached also in the massive gravity context in \cite{D'Amico:2011jj}. In that case however no dynamics where considered for $f_{\mu\nu}$ and the possibility of finding solutions by enforcing the non-dynamical Bianchi constraint was missed due to a too restrictive parameterization of Minkowski space.

If we do proceed and eliminate the source in the above prescribed manner we end up with an evolution equation where the Hubble expansion is driven by terms of exactly the same form as in (\ref{Xconst}). Although it is intriguing that these terms actually do scale in accordance with the usual matter fluid components for large values of $a$, for the purposes of this paper we do not want to impose any such restrictions on the matter sector.

\end{document}